\documentclass[a4paper,10pt]{article}
\usepackage{amsmath,amsfonts}
\usepackage{times}
\usepackage[T1]{fontenc}
\usepackage{epsfig}
\usepackage{amssymb}
\usepackage{color}
\definecolor{NavyBlue}{rgb}{0.1,0.2,1}

\begin{document}

\title{\Huge{\bf Strongly coupled Skyrme-Faddeev-Niemi hopfions}}

\author{C. Adam$^{a)}$\thanks{adam@fpaxp1.usc.es}, 
J. S\'{a}nchez-Guill\'{e}n $^{a)b)}$\thanks{joaquin@fpaxp1.usc.es},  
T. Roma\'{n}czukiewicz$^{c)}$\thanks{trom@th.if.uj.edu.pl},  \\
A. Wereszczy\'{n}ski $^{c)}$\thanks{wereszczynski@th.if.uj.edu.pl}
       \\
       \\ $^{a)}$ Departamento de F\'isica de Part\'iculas, Universidad
       \\ de Santiago, and Instituto Galego de F\'isica de Altas Enerx\'ias
       \\ (IGFAE) E-15782 Santiago de Compostela, Spain
       \\
       \\ $^{b)}$ Sabbatical leave at: Departamento de F\'isica Te\'orica,
       \\ Universidad de Zaragoza, 50009 Zaragoza, Spain
       \\
       \\ $^{c)}$ Institute of Physics,  Jagiellonian University,
       \\ Reymonta 4, Krak\'{o}w, Poland}

\maketitle

\begin{abstract}
The strongly coupled limit of the Skyrme-Faddeev-Niemi model 
(i.e., without quadratic kinetic term) with
a potential is considered on the spacetime $\mathbb{S}^3\times \mathbb{R}$.
For one-vacuum potentials two types of exact Hopf solitons
are obtained. Depending on the value of the Hopf index, we find
compact or non-compact hopfions. The compact hopfions saturate a
Bogomolny bound and lead to a fractional energy-charge formula $E
\sim |Q|^{1/2}$, whereas the non-compact solitons do not saturate the
bound and give $E \sim |Q|$. In the case of potentials with two vacua
compact shell-like hopfions are derived.
\\
Some remarks on the influence of the potential on topological solutions in the
full Skyrme-Faddeev-Niemi model or in (3+1) Minkowski space are
also made.
\end{abstract}

\newpage

\section{Introduction}
The Skyrme-Faddeev-Niemi (SFN) model \cite{fn}, \cite{cho} is a field
theory with hopfions as solitonic excitations. The model is given
by the following Lagrange density
\begin{equation}
L= \alpha (\partial_{\mu} \vec{n})^2 - \beta  [\partial_{\mu}
\vec{n} \times
\partial_{\nu} \vec{n} ]^2 - \lambda V(\vec{n}), \label{baby skyrme}
\end{equation}
where $\vec{n}=(n^1,n^2,n^3)$ is a unit iso-vector living in
$(3+1)$ dimensional Minkowski space-time. Additionally, $\alpha,
\beta, \lambda $ are positive constants. The second term, referred to
as the Skyrme term (strictly speaking the Skyrme term restricted
to $\mathbb{S}^2$) is obligatory in the case of 3 space dimensions
to avoid the Derrick argument for the non-existence of static,
finite energy solutions. The requirement of finiteness of the
energy for static configurations leads to the asymptotic condition
$\vec{n} \rightarrow \vec{n}_0$, as $\vec{x} \rightarrow \infty$,
where $\vec{n}_0$ is a constant vector. Thus, static
configurations are maps $\mathbb{R}^3 \cup \{ \infty \} \cong
\mathbb{S}^3 \rightarrow \mathbb{S}^2$ and therefore can be
classified by the pertinent topological charge, i.e., the Hopf
index $Q\in \pi_3(\mathbb{S}^2) \cong \mathbb{Z}$. Moreover, as
the pre-image of a fixed $\vec{n} \in \mathbb{S}^2$ is isomorphic
to $\mathbb{S}^1$, the position of the core of a soliton
(pre-image of the antipodal point $-\vec{n}_0$) forms a closed, in
general knotted, loop. For a recent detailed review of the SFN model and
related models which support knot solitons we refer to \cite{radu-volkov}.
\\
The physical interest of the SFN model is related to the fact that it may be
applied to several important physical systems. In the context of
condensed matter physics, it has been used to
describe possible knotted solitons for multi-component superconductors
\cite{babaev}, \cite{juha}. In field theory, its importance
originates in the attempts to
relate it to the low energy (non-perturbative), pure
gluonic sector of QCD \cite{fn}, \cite{glueball}.
In this picture, relevant particle excitations,
i.e., glueballs are identified with knotted topological solitons.
This idea is in agreement with the standard picture of mesons,
where quarks are connected by a very thin tube of the gauge field.
Now, because of the fact that glueballs do not consist of quarks,
such a flux-tube cannot end on sources. In order to form a
stable object, the ends must be joined, leading to loop-like
configurations.
\\
Although the SFN model (or some generalization thereof)
might provide the chance for a
very elegant description of the physics of glueballs, this proposal has 
its own problems. First of all, one has to include a symmetry
breaking potential term \cite{niemi2}, although the potential would not
be required for stability reasons. This is necessary in order to avoid the
existence of massless excitations, i.e., Goldstone bosons
appearing as an effect of the spontaneous global symmetry
breaking. Indeed, the Lagrangian without a potential 
possesses global $O(3)$ symmetry while the vacuum state is only
$O(2)$ invariant. Thus, two generators are broken and two massless
bosons emerge. This feature of the SFN model has been
recently discussed and some modifications have been proposed
\cite{niemi2}, \cite{potential}.
\\
Secondly, due to the non-trivial topological as well as geometrical
structure of solitons one is left with numerical solutions only.
The issue of obtaining the global minimum (and local minima) in a
fixed topological sector is a highly complicated, only partially
solved problem (see e.g. \cite{numeric1} and \cite{numeric2} for the case 
without potential).
The interaction between hopfions is, of course, even more
difficult.
\\
In spite of the huge difficulties, some analytical results have
been obtained. One has to underline, however, that they have
been found entirely for the potential-less case. Let us mention
the famous Vakulenko-Kapitansky energy-charge formula, $E \geq c_1
|Q|^{3/4}$ \cite{vk}, \cite{inequal}, \cite{ward}. Similar upper bounds 
$E \leq c_2
|Q|^{3/4}$ have also been
reported \cite{inequal}. Further, interactions in the charge $Q=2$ sector 
have been analyzed and  attractive channels have been reported 
\cite{ward int}. Among analytical approaches which
have been applied to the SFN model, one should mention the generalized
integrability \cite{analytic meth} and the first integration
method \cite{analytic int}, which were especially helpful in
constructing vortex \cite{vortex} and non-topological solutions
\cite{analytic sol}.
\\
Another approach, which sheds some light on the properties of hopfions and
allows for analytical calculations is the substitution of the flat
Minkowski space-time by a more symmetric space as, e.g., $\mathbb{S}^3\times 
\mathbb{R}$
\cite{ward}, \cite{analytic S3}, where an infinite set of static and
time dependent solutions where found.
\\
The main aim of the present paper is to analytically investigate the 
physically important problem of the 
role of the potential term in theories supporting hopfions. It is known 
from other solitonic theories that the inclusion of a potential leads to 
significant changes of geometric as well as dynamical (stability, 
interactions) properties of solitons. Indeed, the influence of the potential 
term on qualitative and quantitative properties of topological solitons has 
been
established in a version of the SFN model in (2+1) dimensions,
i.e., in the baby Skyrme model \cite{karliner}, \cite{baby pot}, 
\cite{double pot}. Further, in the case of the (3+1) dim Skyrme model it 
has been found that the inclusion of the so-called old potential strongly 
modifies the geometrical properties of solitons \cite{massive sk}. However, 
there are almost no results in the case of 
hopfions\footnote{In \cite{numeric1} Gladikowski and Hellmund reported 
on charge 
$Q=1,2$ axial symmetric hopfions in the SFN model with the so-called old 
baby potential.}.   
As we would like to attack the issue analytically, leaving numerics for 
future work, we have to make some simplifications. 
\\
Our strategy will be two-fold: we simplify the action and move to a more 
symmetric base space-time $\mathbb{S}^3\times \mathbb{R}$. 
Specifically, we perform the strong coupling 
$\alpha \rightarrow 0$ limit  \cite{speight}, that is, we neglect the 
quadratic part of the action\footnote{The
limit $\alpha \rightarrow 0$ has been previously investigated in
the context of the baby Skyrme model \cite{GP}, \cite{rest-bS}.}.
This assumption, although leading to a rather peculiar Lagrangian, is 
interesting and quite acceptable because of many reasons. First of all, the
obtained model still allows to circumvent the Derrick arguments against 
the existence of solitonic solutions. The model has also reasonable 
time-dynamics and Hamiltonian formulation as it contains maximally first 
time derivatives squared. This opens the possibility for the collective 
quantization of solitons. Additionally, it explores a class of models 
having, under certain circumstances, BPS  hopfions. The existence of such 
a BPS limit for higher-dimensional topological solitons is a rather 
non-trivial feature  (see \cite{sutcliffe bps}, \cite{skyrme bps} in the 
context of the Skyrme model or \cite{ferreira bps} for the SFN model).  
\\
Moreover, as we comment in the last section, the solution of the 
model in the limit $\alpha \rightarrow 0$ 
probably can be viewed as a zero order
approximation to the true soliton of the full theory. In particular, it 
will be advocated that static properties of hopfions of the SFN model (in 
the assumed curved space) may be qualitatively and quantitatively described 
by solitons of its strongly coupling limit. We find that topological and 
geometrical properties are governed by the strongly coupled model, while 
the kinetic part of the full SFN model only mildly modifies them.  
\\
The second assumption i.e., assuming a non-flat base space, takes us rather far 
from the standard SFN model but it is the price we have to pay
if we want to perform all calculations in an analytical way
while preserving the topological properties. Nonetheless, the
presented results 
may give an intuition and hints about what can happen with true SFN knots on 
$\mathbb{R}^3\times \mathbb{R}$ if the potential term is included.
\section{The strongly coupled Skyrme-Faddeev-Niemi model}

\subsection{The model}
Let us begin with the limit $\alpha \rightarrow 0$ considered above,
leading 
to the following strongly coupled SFN
model 
\begin{equation}
L= - \beta [\partial_{\mu} \vec{n} \times
\partial_{\nu} \vec{n} ]^2 - \lambda V(\vec{n}), \label{b skyrme}
\end{equation}
where the potential is assumed to depend entirely on the third component
$n^3$. 
\\
After the stereographic projection
\begin{equation}
\vec{n}=\frac{1}{1+|u|^2} \left( u+\bar{u}, -i ( u-\bar{u}),
1- |u|^2 \right).
\end{equation}
we get
\begin{equation}
L= - 8 \beta
\frac{(u_{\mu}\bar{u}^{\mu})^2-u_{\mu}^2\bar{u}_{\nu}^2}{(1+|u|^2)^4} -
\lambda V(|u|^2)
\end{equation}
where $u_\mu \equiv \partial_\mu u$, etc.
The corresponding field equations read
\begin{equation}
 \partial_{\mu} \left( \frac{\mathcal{K^{\mu}}}{(1+|u|^2) ^2} \right)+  
\frac{2\bar{u}}{(1+|u|^2)^3} \mathcal{K}_{\mu}
\partial^{\mu} u-
\frac{\lambda}{4} \bar{u} V'=0
\end{equation}
and its complex conjugate. Here prime denotes differentiation with respect 
to $u\bar{u}$ and
\begin{equation}
\mathcal{K}^{\mu}= 4\beta
\frac{(u_{\nu}\bar{u}^{\nu})\bar{u}^{\mu}-\bar{u}_{\nu}^2u^{\mu}}{(1+|u|^2)^2}.
\end{equation}
Thus,
\begin{equation} \label{simp-eq}
 \partial_{\mu} \mathcal{K^{\mu}} -
\frac{\lambda}{4} \bar{u} (1+|u|^2)^2 V'=0,
\end{equation}
where we used the following identity
\begin{equation}
\mathcal{K}^{\mu} \bar{u}_{\mu}=0.
\end{equation}
\subsection{Integrability and area-preserving diffeomorphisms}
Neglecting the standard kinetic part of the SFN action results in an
enhancement of the symmetries of the model.  Indeed, following previous 
works one may easily guess the following infinite family of 
conserved quantities
\begin{equation}
J_{\mu}^G=\frac{\delta G}{\delta \bar{u}} \mathcal{K}_{\mu} - 
\frac{\delta G}{\delta u} \bar{\mathcal{K}}_{\mu},
\end{equation}
where $G=G(u\bar{u})$ is an arbitrary, differentiable function depending 
on the modulus $|u|$. The charges corresponding to the currents are
\begin{equation}
Q^G=\int d^3 x J_0^G
\end{equation}
and obey the abelian subalgebra of area-preserving diffeomorphisms on 
the target space $S^2$ spanned by the complex field $u$ under the Poisson 
bracket, 
\begin{equation}
\{Q^{G_1},Q^{G_2} \}=0. 
\end{equation}
The abelian character of the algebra is enforced by the inclusion of the 
potential term in the action, as the Skyrme term is invariant under the 
full nonabelian algebra of the area-preserving diffeomorphisms on the 
target space $S^2$ \cite{ab-dif}.  
\\
The infinite number of the conserved currents leads to the integrability of the 
model (at least in the sense of the generalized integrability). In fact, 
such a integrable limit of the SFN model has been suggested in \cite{afz}. 
However, because of the fact that the model discussed there
did not contain any potential, this limit gave a theory with unstable solitons. 
\\
Further, one can notice that the existence of the conserved currents does 
not depend on the physical space-time, and therefore is relevant for 
the curved space $\mathbb{S}^3 \times \mathbb{R}$ as well as the flat 
space $\mathbb{R}^3\times \mathbb{R}$.  However, in the case of the 
curved space $\mathbb{S}^3 \times \mathbb{R}$ we will find that
the model reveals a 
very special property. Namely, some of its solutions (the compacton solutions
which are different from the vacuum only on a finite fraction of the base
space $\mathbb{S}^3$) are of BPS type i.e., 
they saturate the pertinent Bogomolny-like inequality between the energy 
and the Hopf charge. Consequently, they obey a first order differential 
equation.  
\\
From a geometrical point of view the strongly coupled model is based on 
the square of the pullback of the volume on the target space. This 
property is shared with the integrable Skyrme model in (2+1) and (3+1) 
dimensions. In contrast to the integrable Skyrme models, here, 
such a term is not the topological charge density squared. Therefore, 
the relation between the Lagrange density and topological current is 
rather obscure, which is one of the reasons why we are not able to 
make more general statements on the conditions for the existence of
BPS type hopfions (i.e., for which base spaces and Ansaetze BPS hopfions
exist). 
What we can say, however, is that BPS type hopfions cannot exist in flat
Minkowski space. The reason is that for a soliton solution which obeys a BPS
equation, the two terms in the lagrangian give equal contributions to the
energy, $E_4 = E_0$ (here $E_4$ is the energy from the term quartic in
derivatives, whereas $E_0$ comes from the potential term with no
derivatives). On the other hand, it easily follows from a Derrick type scaling
argument that in flat space $\mathbb{R}^3$ for any static solution the
energies must obey the virial condition $E_4 = 3 E_0$, which is obviously
incompatible with the BPS condition on the energies for solutions with finite
and non-zero energies. 
\section{Exact solutions on $\mathbb{S}^3 \times 
\mathbb{R} $ }

\subsection{Ansatz and equation of motion}
As mentioned in Section 1, in order to present examples of some exact 
solutions we consider the model on $\mathbb{S}^3 \times \mathbb{R}$, 
where coordinates are chosen such that the
metric is
\begin{equation}
ds^2 = dt^2 -R_0^2\left( \frac{dz^2}{4z(1-z)} +(1-z)d\phi_1^2
+zd\phi_2^2 \right) , \label{metric}
\end{equation}
where $z \in [0,1]$ and the angles $\phi_1, \phi_2 \in [0,2\pi]$,
$R_0$ denotes the radius of $\mathbb{S}^3$.
\\
Moreover, for the moment we choose for the potential  
\begin{equation}
V=\frac{1}{2} (1-n^3). \label{b potential}
\end{equation}
In 2+1 dimensional Minkowski space-time, i.e., in the baby Skyrme model, this
potential is known as the old baby Skyrme potential.
It should be stressed that the fact that the model is solvable does
not depend on the particular form of the potential. However, 
specific quantitative as well as qualitative properties of the
topological solutions are strongly connected with the form of the
potential.
\\
In the subsequent analysis we assume the standard Ansatz
\begin{equation}
u=e^{i(m_1\phi_1+m_2\phi_2)} f(z), \label{ansatz}
\end{equation}
where $m_1, m_2 \in Z$. 
This ansatz exploits the base space symmetries of the theory, which for static
configurations is equal to the isometry group SO(4) of the base space 
$\mathbb{S}^3$. This group has rank two, so it allows the separation of two
angular coordinates $e^{im_l\phi_l}$, $l=1,2$, see e.g. \cite{analytic S3} for
details.
\\
The profile function $f$ can be derived
from the equation
\begin{equation}
-\partial_z \left[ \frac{ f' f^2}{(1+f^2)^2} \Omega \right] +
\left(\frac{ ff'^2}{ (1+f^2)^2} \Omega \right) +
 \tilde{\lambda} f=0,
\end{equation}
where we introduced
\begin{equation}
\Omega= m_1^2z+m_2^2(1-z)
\end{equation}
and
\begin{equation}
\tilde{\lambda}=\frac{\lambda R_0^4}{128 \beta}.
\end{equation}
In order to get a solution with nontrivial topological Hopf charge
one has to impose boundary conditions which guarantee that the
configuration covers the whole $\mathbb{S}^2$ target space at
least once
\begin{equation}
f(z=0)=\infty, \;\;\;\; f(z=1)=0.
\end{equation}
The equation for $f$ can be further simplified leading to
\begin{equation}
f \left(\partial_z \left[ \frac{ f' f}{(1+f^2)^2} \Omega \right] -
 \tilde{\lambda} \right)=0.
\end{equation}
This expression is obeyed by the trivial, vacuum solution $f=0$ or
by a nontrivial configuration satisfying
\begin{equation}
 \partial_z \left[ \frac{ f' f}{(1+f^2)^2} \Omega \right] =
 \tilde{\lambda} \;\; \Rightarrow \;\; \frac{ f' f}{(1+f^2)^2} \Omega =  
\tilde{\lambda}(z+z_0).
\end{equation}
This formula may be also integrated giving finally
\begin{equation}
 \frac{ 1}{1+f^2}  =  -\frac{\tilde{\lambda}}{2}\int dz
 \frac{z+z_0}{m_1^2z+m_2^2(1-z)}+C,
\end{equation}
where $C$ and $z_0$ are real integration constants, whose values
can be found from the assumed boundary conditions.
\\
One can also easily calculate the energy density
\begin{equation}
\varepsilon = \frac{32 \beta}{R_0^4} \frac{4 f^2f'^2}{(1+f^2)^4}
\left( m_1^2z+m_2^2(1-z)  \right)+ \frac{\lambda f^2}{1+f^2}
\end{equation}
and the total energy
\begin{equation}
E=\frac{(2\pi)^2R_0^3}{2} \int_0^1 dz  \varepsilon.
\end{equation}
\subsection{Compact hopfions}
It follows from the results of 
\cite{sign-G}, \cite{GP}, \cite{comp baby}, \cite{rest-bS}
that one should expect the appearance of
compactons in the pure SFN model with the old baby Skyrme potential.
As suggested by its
name, a compacton in flat space is a solution with a finite support, reaching
the vacuum value at a finite distance \cite{comp}. Thus, compactons do
not possess exponential tails but approach the vacuum in a
power-like manner.
On the base space $\mathbb{S}^3$, all solutions are compact (because the base
space itself is compact). By analogy with the flat space case, 
we shall call compactons those
solutions which
are non-trivial (i.e., different from the vacuum) only on a finite fraction of
the base space and join smoothly to the vacuum with smooth first derivative.  
\\
An especially simple situation occurs for the $m_1=\pm m_2 \equiv m$
case. Then, the equation of motion for the profile function reduces to
\begin{equation}
 \partial_z^2 g =\frac{2\tilde{\lambda}}{m^2}, \label{m eom}
\end{equation}
where
\begin{equation}
g =1-\frac{1}{1+f^2}.
\end{equation}
Observe that $g\ge 0$ by the definition of the function $g$. 
The pertinent boundary conditions for compact hopfions are
$f(0)=\infty$ and $f(z=z_R)=0$, where $z_R \leq 1$ is the radius
of the compacton. In addition, as one wants to deal with a globally
defined solution, the compact hopfion must be glued with the
trivial vacuum configuration at $z_R$, i.e., $f'(z=z_R)=0$. In
terms of the function $g$ we have $g(0)=1$, $g(z=z_R)=0$ and
$g_z(z=z_R)=0$. Thus, the compacton solution is
\begin{equation}
g(z) = \left\{
\begin{array}{lc}
\left(1-\frac{z\sqrt{\tilde{\lambda}} }{m} \right)^2 & z \leq z_R
\\
0 & z \geq z_R.
\end{array} \right.
\end{equation}
We remark that the energy density in terms of the function $g$ 
and for $m_1 = m_2 =m$ may be expressed
like
\begin{equation}
\varepsilon = \frac{128 \beta}{R_0^4} \left( \frac{m^2}{4} g'^2
+ \tilde \lambda g \right) 
\end{equation}
which makes it obvious that the vacuum configuration $g\equiv 0$ minimizes the
energy functional.
The size of the compact soliton is
$$z_R=\frac{m}{\sqrt{\tilde{\lambda}}}.$$
As the $z$ coordinate is restricted to the interval $[0,1]$, 
we get a limit for the topological charge for possible compact
solitons. Namely
\begin{equation}
m \leq \sqrt{\tilde{\lambda}}= \frac{\sqrt{\lambda} R_0^2}{\sqrt{128
\beta}}.
\end{equation}
In other words, one can derive a compact hopfion solution provided
that its topological charge does not exceed a maximal value
$Q_{max}=\lfloor \tilde{\lambda} \rfloor$, which is fixed once
$\lambda, \beta, R_0$ are given.
\\
Further, the energy density onshell is
\begin{equation}
\varepsilon = 2\lambda g
\end{equation}
and the total energy
\begin{equation}
E=(2\pi)^2 \lambda R_0^3 \int_0^{\frac{m}{\sqrt{\tilde{\lambda}}}}
dz \left(1-\frac{z\sqrt{\tilde{\lambda}} }{m} \right)^2=(2\pi)^2
\lambda R_0^3 \frac{m}{\sqrt{\tilde{\lambda}}}
\frac{1}{3}=\frac{32 \sqrt{2}\pi^2}{3}  \sqrt{\lambda \beta} m
R_0.
\end{equation}
Taking into account the expression for the Hopf index $$Q=m_1 m_2
= m^2.$$ we get
\begin{equation}
E=\frac{32 \sqrt{2}\pi^2}{3}  \sqrt{\lambda \beta} R_0 \;\;
|Q|^{\frac{1}{2}}, \;\;\;\; |Q| \leq |Q_{max}|.
\end{equation}
For a generic situation, when $m_1^2 \neq m_2^2$, we find the
exact solutions
\begin{equation}
 g(z)=1+\frac{2\tilde{\lambda}}{m_1^2-m_2^2} \left[ z - \left(z_R + 
\frac{m_2^2}{m_1^2-m_2^2} \right) \ln \left( 1+
 z\frac{m_1^2-m_2^2}{m_2^2}\right)\right].
\end{equation}
In this case, the size of the compacton $z_R$ is given by a solution of
the non-algebraic equation
\begin{equation}
z_R - \left(z_R + \frac{m_2^2}{m_1^2-m_2^2} \right) \ln \left( 1+
 z_R\frac{m_1^2-m_2^2}{m_2^2}\right)+\frac{m_1^2-m_2^2}{2\tilde{\lambda}}=0.
 \label{zR}
\end{equation}
\subsection{Non-compact hopfions}
Let us again consider the profile function equation for $m_1=\pm
m_2$ (\ref{m eom}) but with non-compacton boundary
conditions. Namely, $g(0)=1$, $g(z=1)=0$, i.e., the solutions
nontrivially cover the whole $\mathbb{S}^3$ base space. The
pertinent solution reads
\begin{equation}
g (z)= \frac{\tilde{\lambda}}{m^2} z^2
-\left(1+\frac{\tilde{\lambda}}{m^2} \right)z+1.
\end{equation}
However, this solution makes sense only if the image of $g$ is
not negative. This is the case if
\begin{equation}
\frac{\tilde{\lambda}}{m^2} \leq 1 \;\; \Rightarrow \;\; m \geq
\sqrt{\tilde{\lambda}}
\end{equation}
and we found a lower limit for the Hopf charge. Thus, such
non-compact hopfions occur if their topological charge is larger
than a minimal charge $Q_{min} = \lceil \tilde{\lambda} \rceil$.
\\
The corresponding energy is
\begin{equation}
E=\frac{(2\pi)^2}{2} \lambda R_0^3 \;\; \left[
\frac{32\beta}{R_0^4} |Q| \left( 1- \frac{\lambda R_0^4}{128 \beta
|Q|}\right)^2 + \lambda \left( 1 -\frac{1}{3} \frac{R_0^4
\lambda}{128 \beta |Q|}\right) \right],
\end{equation}
for $ |Q| \geq |Q_{min}|$.
\\
Finally we are able to write down a formula for the total energy
for a soliton solution with a topological charge $Q$
\begin{equation}
E=\left\{
\begin{array}{lc}
\frac{32 \sqrt{2}\pi^2}{3}  \sqrt{\lambda \beta} R_0 \;\;
|Q|^{\frac{1}{2}}& |Q| \leq \lfloor \frac{\lambda R_0^4}{128 \beta} \rfloor 
\\ &  \\
\frac{(2\pi)^2}{2} \lambda R_0^3 \;\; \left[ \frac{32\beta}{R_0^4}
|Q| \left( 1- \frac{\lambda R_0^4}{128 \beta |Q|}\right)^2 +
\lambda \left( 1 -\frac{1}{3} \frac{R_0^4 \lambda}{128 \beta
|Q|}\right) \right] & |Q| \geq \lceil\frac{\lambda R_0^4}{128
\beta}\rceil,
\end{array} \right.
\end{equation}
where the first line describes the compact hopfions and the second
one the standard non-compact solitons.

\vspace*{0.3cm}

\noindent {\bf Remark:} The pure Skyrme-Faddeev-Niemi model with
potential (\ref{b potential}) can be mapped, after the dimension
reduction, on the signum-Gordon model \cite{sign-G}.
\\
Indeed, if we rewrite the energy functional using our Ansatz with
$m_1=\pm m_2$, and take into account the definition of the
function $g$, then we get the energy for the real signum-Gordon
model
\begin{equation}
E=\frac{(2\pi)^2R_0^3}{2} \int_0^1dz \left( \frac{32\beta
m^2}{R_0^4}g_z^2 +\lambda g  \right).
\end{equation}
The signum-Gordon model is well-known to support compact solutions, so this
map is one simple way to understand their existence. The same is true on
two-dimensional Euclidean base space, explaining the existence of compactons
in the model of Ref. \cite{GP} (to our knowledge, compactons in a relativistic
field theory have been first discussed in that reference).

\vspace*{0.3cm}

\noindent {\bf Remark:} Compact hopfions saturate the BPS bound,
whereas non-compact hopfions do not saturate it.
\\
This follows immediately from the last expression and the fact that
all solitons are solutions of a first order ordinary differential
equation. Namely,
\begin{equation} \label{BPS-energy}
E=\frac{(2\pi)^2R_0^3}{2} \int_0^1dz \left[ \left(
\sqrt{\frac{32\beta m^2}{R_0^4}} g_z + \sqrt{\lambda} g^{1/2}
\right)^2 -2\sqrt{\frac{32\beta m^2}{R_0^4}} g_z \sqrt{\lambda}
g^{1/2}\right].
\end{equation}
Then,
\begin{equation}
E \geq -2 \frac{(2\pi)^2R_0^3}{2}\sqrt{\frac{32\beta \lambda
m^2}{R_0^4}}
 \int_{g(0)}^{g(z_R)} dz g_z g^{1/2}
\end{equation}
and
\begin{equation}
E \geq \frac{32\sqrt{2} \pi^2}{3}\sqrt{\lambda \beta}R_0
(g(0)^{3/2}-g(z_R)^{3/2})=\frac{32\sqrt{2} \pi^2}{3}\sqrt{\lambda
\beta}R_0,
\end{equation}
as $g(0)=1$ and $g(z_R)=0$. The inequality is saturated if the
first term in Eq. (\ref{BPS-energy}) vanishes i.e.,
\begin{equation}
\frac{32\beta m^2}{R_0^4} g_z^2 = \lambda g,
\end{equation}
which is exactly the first order equation obeyed by the compact
hopfions. On the other hand, the non-compact solitons satisfy
\begin{equation}
\frac{32\beta m^2}{R_0^4} g_z^2 = \lambda g +C,
\end{equation}
where $C$ is a non-zero constant
$$C=\left(1-\frac{\tilde{\lambda}}{m^2}\right)^2.
$$
\subsection{More general potentials}
The generalization to the models with the
potentials
\begin{equation}
V_s=\lambda \left( \frac{1}{2} (1-n^3)\right)^s, \label{s
potential}
\end{equation}
where $s \in (0,2)$ leads to similar compact solutions. Namely,
\begin{equation}
g(z) = \left\{
\begin{array}{lc}
\left(1-\frac{z\sqrt{\tilde{\lambda}} (2-s) }{m}
\right)^{\frac{2}{2-s}} & z\leq z_R \\
0 & \geq z_R.
\end{array} \right.
\end{equation}
Now, the size of the compacton is
\begin{equation}
z_R= \frac{m}{z\sqrt{\tilde{\lambda}} (2-s)},
\end{equation}
and the limit for the maximal allowed topological charge (in the $m_1=\pm
m_2$ case) is
\begin{equation}
m \leq   \sqrt{\tilde{\lambda}} (2-s).
\end{equation}
For a larger value of the Hopf index one gets a non-compact
hopfion. The energy-charge relation remains (up to a
multiplicative constant) unchanged.
\\
In the limit when $s=2$, i.e.,
\begin{equation}
V_2=\lambda \left( \frac{1}{2} (1-n^3)\right)^2, \label{2
potential}
\end{equation}
we get only non-compact hopfions
\begin{equation}
g (z)=\cosh \left(\frac{2z\sqrt{\tilde{\lambda}}}{m} \right)-
\coth \left(\frac{2\sqrt{\tilde{\lambda}}}{m} \right) \sinh
\left(\frac{2z\sqrt{\tilde{\lambda}}}{m} \right).
\end{equation}
The total energy is found to be
\begin{equation}
E=\frac{(2\pi)^2}{2} \lambda R_0^3   \frac{m}{4
\sqrt{\tilde{\lambda}}} \left( \coth
\frac{2\sqrt{\tilde{\lambda}}}{m}+
  \frac{\frac{2\sqrt{\tilde{\lambda}}}{m}}{\sinh^2 \left(
\frac{2\sqrt{\tilde{\lambda}}}{m} \right)}\right) .
\end{equation}
Asymptotically, for large topological charge $Q=\pm m^2$ we get
\begin{equation}
E=\frac{(2\pi)^2}{2} \lambda R_0^3  \left( \frac{128\beta}{\lambda
R_0^4}|Q| + \frac{1}{45} \frac{\lambda R_0^4}{32 \beta |Q|}
\right).
\end{equation}
Finally, let us comment that for $s > 2$ there are no
finite energy compact hopfions, at least as long as the Ansatz is assumed.
Indeed, the Bogomolny equation for $g$ in this case is
$$
g_z^2 = \frac{4\tilde \lambda}{m^2}g^s
$$
and the power-like approach to the vacuum $g \sim (z-z_R )^\alpha $ leads to
$$
\alpha = \frac{2}{2-s}
$$
which is negative for $s>2$. There may, however, exist non-compact hopfions.
In the case $s=4$, for instance
(the so-called holomorphic potential in the baby
Skyrme model), the resulting first order equation for $g$ is
$$
g_z^2 = \frac{4\tilde \lambda}{m^2}(g^4 + g_0^4)
$$
the general solution of which is given by
the elliptic integral
$$
\int_{g=0}^{g=g(z)} \frac{dg}{(g^4+g_0^4)^{1/2}} =
- \frac{2}{|m|}\sqrt{\tilde\lambda} (z-z_0)
$$
(we chose the negative sign of the root because $g$ is a decreasing function 
of $z$), and we have to impose the boundary conditions
$$ 
g(z=1)=0 \quad \Rightarrow z_0=1
$$
and $g(z=0)=1$ which leads to
$$
\int_{0}^{1} \frac{dg}{(g^4+g_0^4)^{1/2}} =
 \frac{2}{|m|}\sqrt{\tilde\lambda} .
$$
The last condition can always be fulfilled because the l.h.s. becomes
arbitrarily large for sufficiently small values of $g_0$ and vice versa.

\subsection{Double vacuum potential}
Another popular potential often considered in the context of the
baby skyrmions, and referred to as the new baby Skyrme potential, is
given by the following expression
\begin{equation}
V=1-(n^3)^2. \label{pot double}
\end{equation}
In contrast to the cases considered before, this potential has two vacua at
$n^3 = \pm 1$.
After taking into account
the Ansatz and the definition of the function $g$, the equation of motion reads
\begin{equation}
\frac{1}{2} \partial_z (\Omega g_z) = \tilde{\lambda}4(1-2g),
\end{equation}
leading, for $m_1=\pm m_2$, to the general solution
\begin{equation}
g(z)=\frac{1}{2}\left( 1- \sqrt{1+4C} \sin \left(
\frac{4\sqrt{\lambda} (z-z_0)}{m}\right) \right),
\end{equation}
where $C, z_0$ are constants.
\\
Here, we start with the non-compact solitons. Then, assuming the
relevant boundary conditions we find
\begin{equation}
g(z)=\frac{1}{2} \left[1- \frac{\sin \frac{4\sqrt{\lambda} }{m}
(z-\frac{1}{2}) }{\sin \frac{2\sqrt{\tilde{\lambda}}}{m} }
\right].
\end{equation}
This configuration describes a single soliton if $g$ is a
monotonous function from 1 to 0. This implies that the sine has to be
a single-valued function on the interval $z\in [0,1]$, i.e.,
\begin{equation}
\frac{4\sqrt{\tilde{\lambda}}}{m} \leq \pi \;\;\;\Rightarrow
\;\;\; |Q|\geq \frac{16 \tilde{\lambda}}{\pi^2}.
\end{equation}
Exactly as before, the non-compact solutions do not saturate the
corresponding Bogomolny bound.
\\
For a sufficiently small value of the topological charge we obtain a
one-parameter family of compact hopfions
\begin{equation}
g(z)= \left\{
\begin{array}{cc}
1 & 0 \leq z \leq z_r\\
\frac{1}{2} \left[1-  \sin \frac{4\sqrt{\lambda} }{m} (z-z_0)
\right] & z_r
\leq z \leq z_R \\
0 & z \geq z_R
\end{array} \right.,
\end{equation}
where the boundary conditions have been specified as $g(z_r)=1,
g(z_R)=0$ and $ g'(z_r)=g'(z_R)=0$. The inner and outer boundaries  of
the compacton are located at
\begin{equation}
z_r=z_0+\frac{\pi m }{8\sqrt{\tilde{\lambda}}}, \;\;\;
z_R=z_0+\frac{3\pi m }{8\sqrt{\tilde{\lambda}}}
\end{equation}
and $z_0$ is a free parameter restricted to
\begin{equation}
z_0 \in [-\frac{\pi m }{8\sqrt{\tilde{\lambda}}} , 1-\frac{3\pi m
}{8\sqrt{\tilde{\lambda}}}].
\end{equation}
We remark that in this case the energy density in terms of the function 
$g$ may be expressed
like
\begin{equation}
\varepsilon = \frac{128 \beta}{R_0^4} \left( \frac{1}{4} g'^2
+ \tilde \lambda g (1-g) \right) 
\end{equation}
which makes it obvious again that both vacuum configurations 
$g =0,1$ minimize the
energy functional. 
\\
As we see, compact solutions in the model with the new baby Skyrme
potential are shell-like objects. In fact, there is a striking
qualitative resemblance between the baby skyrmions and the compact
hopfions in the pure Skyrme-Faddeev-Niemi model with potentials
(\ref{b potential}), (\ref{pot double}). Namely, it has been
observed that the old baby skyrmions are rather standard solitons
with or without rotational symmetry, whereas the new baby
skyrmions possess a ring-like structure \cite{double pot}. 
Here, in the case of the
new baby potential, we get a higher dimensional generalization of
ring structures, i.e., shells.
\\
The energy-charge relation again takes the form of the square root
dependence for compactons,
\begin{equation}
E=\frac{\pi^3}{2} R_0 \sqrt{128 \beta \lambda} \;\; |Q|^{1/2},
\end{equation}
where we used the fact that the compact solutions saturate the
Bogomolny bound.
\\
{\bf Remark:} Observe that one may construct an onion 
type structure of non-interacting shell hopfions with a total energy 
which goes linearly with the total charge. 
When these hopfions are sufficiently separated they form a meta-stable 
solution, but the total energy of a single hopfion ring with the same 
total charge is smaller (it goes like $\sqrt{Q}$). 
Therefore, one may expect that the onion solution is not stable.
\subsection{Free model case}
To have a better understanding of the role of the potential let us briefly
consider the case without potential, i.e., $\lambda=0$. In this
case one can easily find the hopfions \cite{analytic S3}
\begin{equation}
g(z)=1-\frac{\ln \left(1+z\frac{m_1^2-m_2^2}{m_2^2}\right)}{\ln
\left( 1+\frac{m_1^2-m_2^2}{m_2^2}\right)}
\end{equation}
for $m_1^2 \neq m_2^2$ and
\begin{equation}
g(z)=1-z
\end{equation}
for $m_1=\pm m_2$. As we see, all solitons are of the non-compact
type, which differs profoundly from the previous situation. \\
The energy-charge formula reads
\begin{equation}
E= \frac{(2\pi)^2 \beta }{4R_0}\; \frac{m_1^2-m_2^2}{\ln m_1 -
\ln m_2}
\end{equation}
or for $m_1^2 = m_2^2$
\begin{equation}
E= \frac{(2\pi)^2 \beta }{2R_0} |Q|.
\end{equation}
Again, the difference is quite big as we re-derived the standard
linear dependence.
\\
{\bf Remark:} There exists a significant difference between models which have
the quartic, pure Skyrme term as the only kinetic term (containing
derivatives) on the one hand, and models which have a standard quadratic
kinetic term (either in addition to or instead of the quartic Skyrme term), on
the other hand. Models with a quadratic kinetic term have the typical vortex
type behaviour 
$$
u \sim r^m e^{im\phi} 
$$ 
near the zeros of $u$. Here $r$ is a generic radial variable, $\phi $ is a
generic angular variable wrapping around the zero, and $m$ is the winding
number. In other words, configurations with higher winding about a zero of $u$
are higher powers of the basic $u$ with winding number one, where both the
modulus and the phase part of $u$ are taken to a higher power. This behaviour
is, in fact, required by the finiteness of the Laplacian $\Delta u$ at $r=0$.
Models with only a quartic pure Skyrme kinetic term (both with and without
potential), however, show the
behaviour
$$
u \sim r e^{im\phi} 
$$  
i.e., only the phase is taken to a higher power for higher winding. For our
concrete model on base space $\mathbb{S}^3$, and for the simpler case $m_1 =
m_2 \equiv m$, we have $u \sim z^{-1/2} e^{im(\phi_1 + \phi_2)}$ near $z=0$
(both with and without a potential term), but with the help of the symmetries
$u \to (1/u)$ and $u \to \bar u$ this may be brought easily to the form 
$$
u \sim \sqrt{z} e^{im(\phi_1 + \phi_2)}, 
$$
as above. As said, the Laplacian acting on this field is singular at $z=0$, so
the field has a conical singularity at this point. One may wonder whether this
singularity shows up in the field equation and requires the introduction of a
delta-like source term. The answer to this question is no. Thanks to the
specific form of the quartic kinetic term, the second derivatives in the field
equation show up in such a combination that the singularity cancels and the
field equation is well-defined at the zero of $u$. As this behaviour is
generic and only depends on the Skyrme term and on the existence of
topological solutions (and not on the base space) we show it for the simplest
case with base space $\mathbb{R}^2$ (i.e., the model of Gisiger and
Paranjape), where $r$ and $\phi$ are just polar coordinates in this space. A
compact soliton centered about the origin behaves like $ u \sim r e^{im\phi} $
near the origin, and has the singular Laplacian
$$ \Delta u = (1-m^2) r^{-1} e^{-m\phi}. $$
On the other hand, the field equation (\ref{simp-eq}) is finite at $r=0$,
because the vector $\vec {\cal K}$ behaves like
$$
\vec {\cal K} = 8\beta \frac{m^2 \hat e_r -im\hat e_\phi }{(1+r^2)^2}
e^{-im\phi} \equiv {\cal K}_r \hat e_r + {\cal K}_\phi \hat e_\phi
$$ 
(here $\hat e_r$ and $\hat e_\phi$ are the unit vectors along the
corresponding coordinates),
and its divergence (which enters into the field equation) is
$$\nabla \cdot \vec {\cal K} \equiv \frac{1}{r} \partial_r (r {\cal K}_r) +
\frac{1}{r} \partial_\phi {\cal K}_\phi = 
\frac{32\beta r}{(1+r^2)^3} e^{-im\phi}
$$
and a potential singular $(1/r)$ contribution cancels between the first and
the second term. As said, this behaviour is completely generic for models with
the Skyrme term as the only kinetic term. These fields, therefore,
solve the field equations also at the singular points $u=0$ and are, 
consequently, strong solutions of the corresponding variational problem.  

{\bf Remark:} In Section 5 we compare numerical solutions of the full model
with the corresponding exact solutions of the strongly coupled model. We shall
find that these concrete results precisely confirm the conclusions of the above
discussion. 
\section{Compact strings in Minkowski space}
In the (3+1) dimensional standard Minkowski space-time we are
not able to find analytic soliton solutions with finite energy, 
because the symmetries of the model do not allow for a symmetry
reduction to an ordinary differential equation in this case.
We may, however,
derive static and time-dependent solutions with a compact
string geometry with the string oriented, e.g. along the $z$ direction. 
These strings have finite energy per unit length in the $z$ direction.
Further, the pertinent topological charge is the
winding number $Q=n$. In this section $(x,y,z)$ refer to the standard
cartesian coordinates in flat Euclidean space. Further, we use the old baby
Skyrme potential of Section 2.1.
\\
The Ansatz we use reads
\begin{equation}
u=f(r)e^{in\phi} e^{i(\omega t +kz)},
\end{equation}
where $\omega, k$ are real parameters, $r^2 \equiv x^2 + y^2$, 
$\phi = \arctan (y/x)$, and $n$ fixes the
topological content of the configuration. It gives the following
equation for the profile function $f$
\begin{equation}
f \left(\frac{1}{r} \partial_r \left[ r\frac{ f' f}{(1+f^2)^2}
\Omega \right] -
 \tilde{\lambda} \right)=0,
\end{equation}
where $\tilde{\lambda}= \lambda/32 \beta$ and
\begin{equation}
\Omega=k^2-\omega^2+\frac{n^2}{r^2}.
\end{equation}
The simplest solutions may be obtained for $\omega^2 = k^2$. Then,
after introducing
\begin{equation}
x= \frac{r^2}{2}, \;\;\; \mbox{and} \;\;\; g=1-\frac{1}{1+f^2}
\end{equation}
we get
\begin{equation}
g_{xx}=\frac{2\tilde{\lambda}}{ n^2}.
\end{equation}
The compact solution reads
\begin{equation}
g(r)=\left\{
\begin{array}{lc}
 \left(1-r^2\frac{\sqrt{\tilde{\lambda}}}{n\sqrt{2}}
\right)^2 & r \leq \frac{\sqrt{n}\sqrt[4]{2}}{\sqrt[4]{\tilde{\lambda}}} \\
0 & r \geq \frac{\sqrt{n}\sqrt[4]{2}}{\sqrt[4]{\tilde{\lambda}}} .
\end{array} \right.
\end{equation}
The total energy (per unit length in $z$-direction) is
\begin{equation}
E=\int d^2 x \frac{8\beta}{(1+|u|^2)^4} [(\nabla u \nabla
\bar{u})^2-(\nabla u)^2 (\nabla \bar{u})^2]
\end{equation}
\begin{equation}
 +
\frac{8\beta}{(1+|u|^2)^4} [ 2u_0\bar{u}_0 (\nabla u \nabla
\bar{u}) - u_0^2 (\nabla \bar{u})^2 -\bar{u}_0^2 (\nabla u)^2] +
\lambda \frac{|u|^2}{ 1+|u|^2 },
\end{equation}
or after inserting our Ansatz
\begin{equation}
 E=2\pi \int_0^{\infty} r dr  \left(  \frac{32
\beta f^2f'^2}{(1+f^2)^4} \left(\frac{n^2}{r^2}+\omega^2
+k^2\right) + \frac{\lambda f^2}{ 1+f^2 } \right)
\end{equation}
and finally
\begin{equation}
E=\frac{2\pi}{3} \left[12 \sqrt{\lambda \beta} |Q|+32 \beta
\omega^2 \right].
\end{equation}
A more complicated case is for $\delta^2 \equiv k^2-\omega^2 >0$.
Then, $\Omega=\delta^2+\frac{n^2}{r^2}$, and the equation for $g$ is
\begin{equation}
\partial_x \left( g_x (n^2+2\delta^2 x)  \right)
-2\tilde{\lambda}=0.
\end{equation}
The compacton solution (with the compacton boundary conditions) is
\begin{equation}
g(x)=1+ \frac{\tilde{\lambda}}{\delta^2} \left[ x -
(\frac{n^2}{2\delta^2} + x_R) \ln \left(1+\frac{2\delta^2 x }{n^2}
\right)\right],
\end{equation}
where $x_R$ is given by
\begin{equation}
1+ \frac{\tilde{\lambda}}{\delta^2} \left[ x_R -
(\frac{n^2}{2\delta^2} + x_R) \ln \left(1+\frac{2\delta^2 x_R
}{n^2} \right)\right]=0.
\end{equation}
\section{The full Skyrme-Faddeev-Niemi model on 
$\mathbb{S}^3\times \mathbb{R}$}
Here, we want to study the relation between solitons of the full SFN model 
and its strongly coupled version.  
Concretely, we assume the old 
baby potential. Then, the full SFN model reads
\begin{equation}
L_{SFN}= 4\alpha \frac{u_{\mu}\bar{u}^{\mu}}{(1+|u|^2)^2} - 8 \beta
\frac{(u_{\mu}\bar{u}^{\mu})^2-u_{\mu}^2\bar{u}_{\nu}^2}{(1+|u|^2)^4} -
\lambda \frac{|u|^2}{1+|u|^2}
\end{equation}
Firstly, let us remark that the symmetric ansatz (\ref{ansatz}) works 
for the full SFN model 
on $\mathbb{S}^3$, although it should be noticed that the energy 
minima obtained within this ansatz do not have to be global minima of 
the model in a fixed topological sector. In fact, to get true minima 
one is forced to solve a 3D numerical problem, which seems to 
be as complicated as in the case of $\mathbb{R}^3$ space. Nonetheless, 
symmetric configurations give an upper bound for true energies, and this
is enough for our purposes, because we mainly want to understand the limiting
case $\alpha \to 0$. 

\begin{figure}[h!]
\begin{center}
\includegraphics[angle=-90.0,width=1.0 \textwidth]{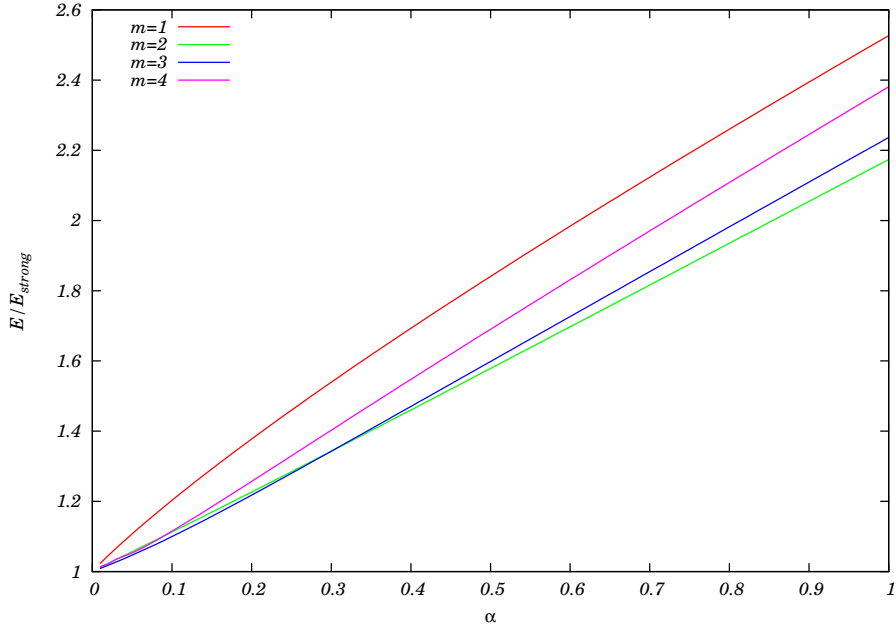}
\caption{Comparison of the energy in the full and the strongly coupled 
(BPS) models as a function of the coupling constant $\alpha$,
for Hopf charge $Q=m^2 = 1,4,9,16$. 
The fixed parameter values are $R_0=5$, $\beta = 0.25$ and $\lambda =1$.}
\label{fig1}
\end{center}
\end{figure}

\begin{figure}[h!]
\begin{center}
\includegraphics[angle=-90.0,width=1.0 \textwidth]{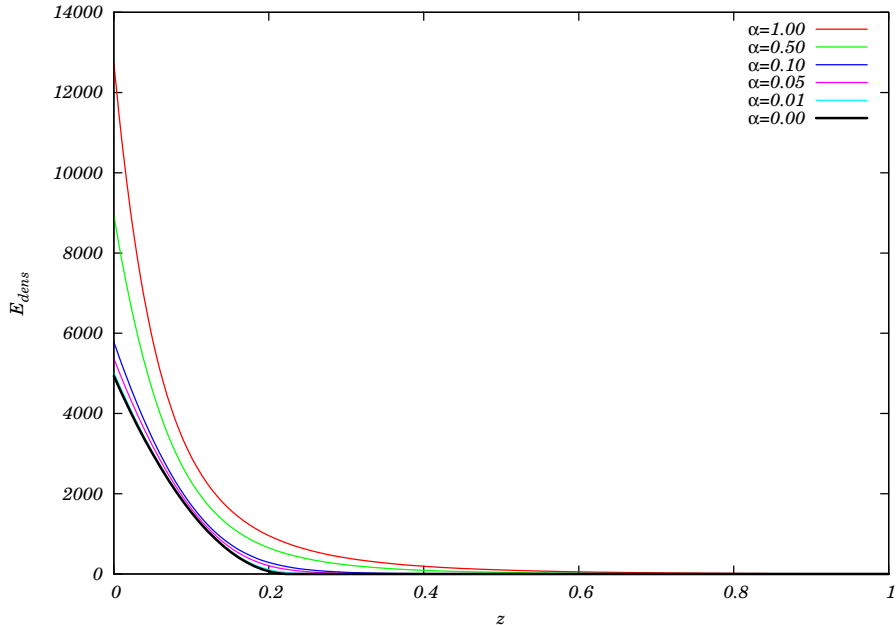}
\caption{Energy densities in the full model, for different values of 
$\alpha$, and in the strongly coupled 
(BPS) model ($\alpha =0$) for Hopf charge $Q=m^2 = 1$, as a
function of $z$.
The fixed parameter values are $R_0=5$, $\beta = 0.25$ and $\lambda =1$.}
\label{fig2}
\end{center}
\end{figure}

\begin{figure}[h!]
\begin{center}
\includegraphics[angle=-90.0,width=1.0 \textwidth]{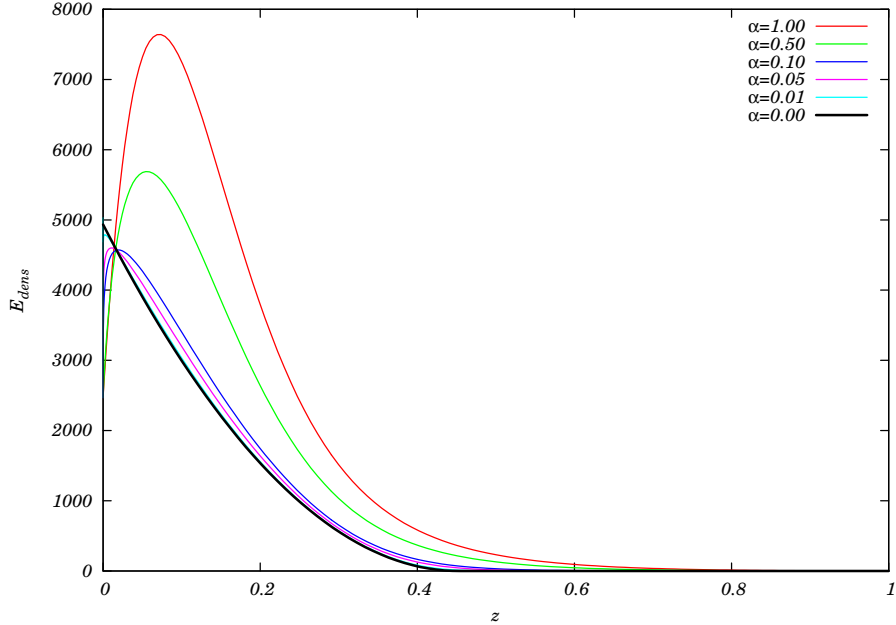}
\caption{Energy densities in the full model, for different values of 
$\alpha$, and in the strongly coupled 
(BPS) model ($\alpha =0$) for Hopf charge $Q=m^2 = 4$, as a
function of $z$.
The fixed parameter values are $R_0=5$, $\beta = 0.25$ and $\lambda =1$.}
\label{fig3}
\end{center}
\end{figure}

\begin{figure}[h!]
\begin{center}
\includegraphics[angle=-90.0,width=1.0 \textwidth]{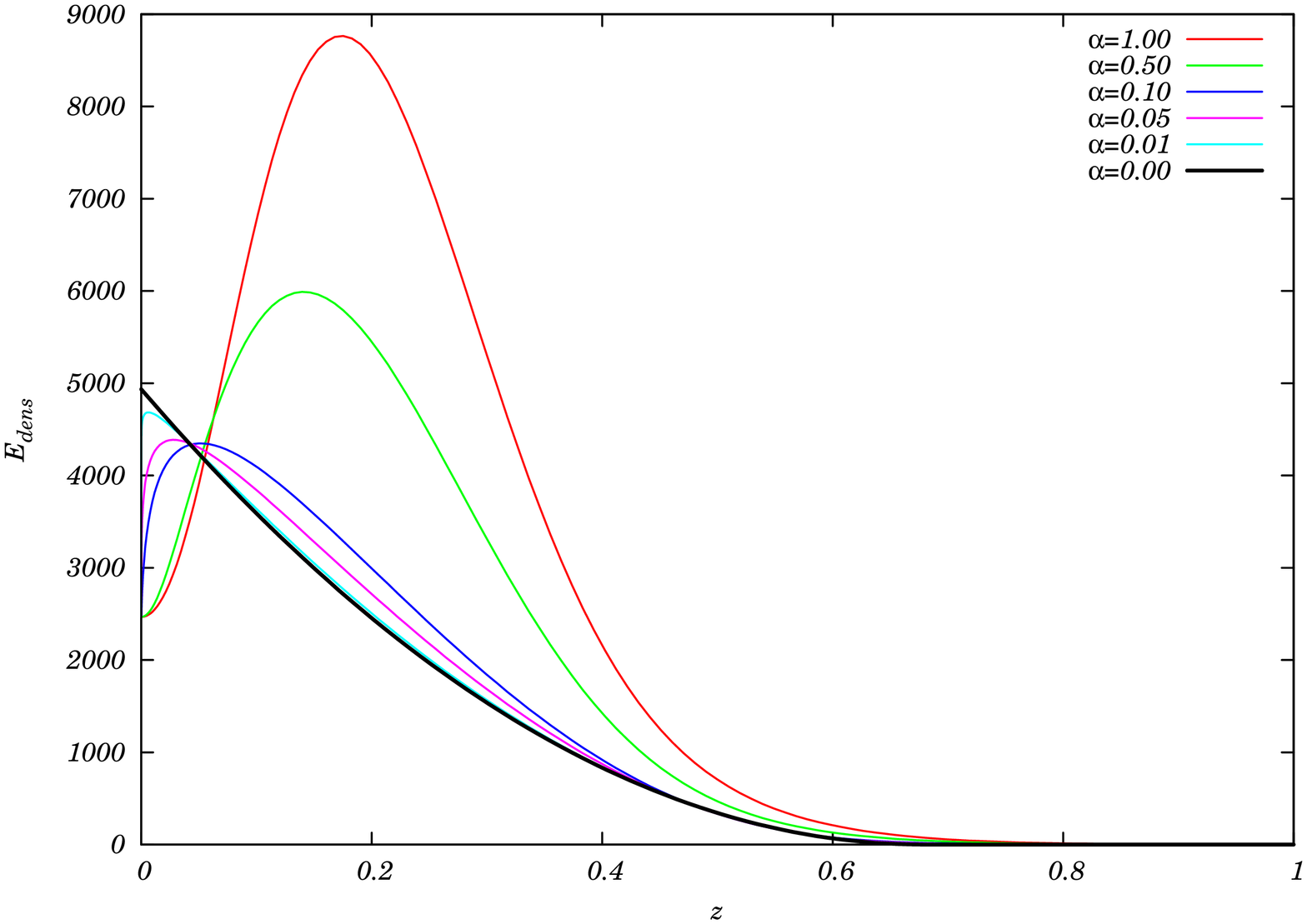}
\caption{Energy densities in the full model, for different values of 
$\alpha$, and in the strongly coupled 
(BPS) model ($\alpha =0$) for Hopf charge $Q=m^2 = 9$, as a
function of $z$.
The fixed parameter values are $R_0=5$, $\beta = 0.25$ and $\lambda =1$.}
\label{fig4}
\end{center}
\end{figure}

\begin{figure}[h!]
\begin{center}
\includegraphics[angle=-90.0,width=1.0 \textwidth]{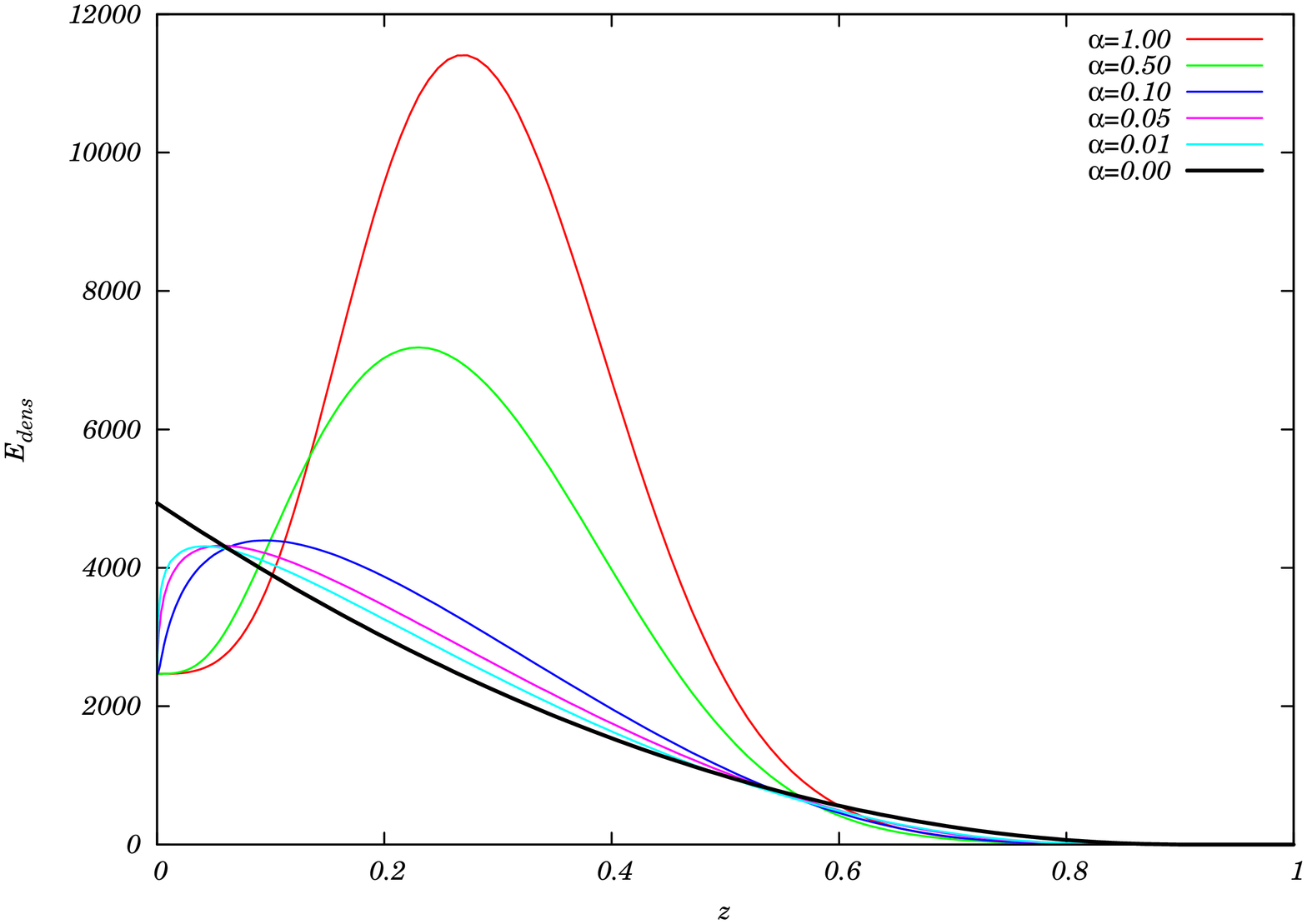}
\caption{Energy densities in the full model, for different values of 
$\alpha$, and in the strongly coupled 
(BPS) model ($\alpha =0$) for Hopf charge $Q=m^2 = 16$, as a
function of $z$.
The fixed parameter values are $R_0=5$, $\beta = 0.25$ and $\lambda =1$.}
\label{fig5}
\end{center}
\end{figure}

The pertinent equation for the profile function 
reads
\begin{equation}
\frac{4\alpha}{R_0^2} \left[ 4\partial_z(z(1-z)f') - 
\frac{f \Omega}{z(1-z)} \right] -\frac{8\alpha}{R_0^2} \frac{f}{1+f^2} 
\left[4z(1-z)f'^2 - \frac{f^2 \Omega}{z(1-z)} \right]
\end{equation}
\begin{equation}
+\frac{128\beta}{R_0^4} \left[ \partial_z \left( 
\frac{\Omega f'f^2}{(1+f^2)^2} \right) - \frac{\Omega ff'^2}{(1+f^2)^2}
\right] - \lambda f=0.
\end{equation}
We solve this equation numerically and then determine the resulting energy and
energy density (in $z$), which may be read off from the energy expression 
\begin{equation}
E=\frac{(2\pi)^2R_0^3}{2} \int_0^1 dz\left(  \frac{4 \alpha}{R_0^2} 
\frac{4z(1-z)f'^2+\Omega f^2}{(1+f^2)^2} +  \frac{32 \beta}{R_0^4} 
\frac{4 \Omega f^2f'^2}{(1+f^2)^4} + \frac{\lambda f^2}{1+f^2} \right).
\end{equation}
In Figure \ref{fig1}, we plot the ratio of the (numerically calculated) 
energy of the full model to the (analytically determined) energy of the
strongly coupled model, for topological charges $Q=m^2 =1,4,9,16$. We find
that in the limit $\alpha \to 0$, the ratio tends to one, for all values of
the topological charge.  
In a next step, we compare the corresponding energy densities. Here, we find a
different behaviour for $m=1$, on the one hand, and for $|m|>1$, on the other
hand. In Figure \ref{fig2}, we compare the (numerical)
energy densities for $m=1$ for
different values of $\alpha$ with the (analytical) energy density for $\alpha
=0$ (strongly coupled model). We find that the energy density for small
$\alpha$ uniformly approaches the $\alpha =0$ curve in the whole interval
$z\in [0,1]$. 
In Figures \ref{fig3} - \ref{fig5}, we compare the (numerical)
energy densities for $m=2,3,4$ for
different values of $\alpha$ with the (analytical) energy density for $\alpha
=0$ (strongly coupled model). In this case, we find that the curves for small
but nonzero $\alpha$ approach the curve for $\alpha =0$ almost
everywhere. There remains, however, a difference near $z=0$, where the curves
for non-zero $\alpha$ approach a different value than the energy density for 
$\alpha =0$. 
The value at $z=0$ for non-zero $\alpha$ is, in fact, just one-half of the 
value for the case $\alpha =0$, as follows easily from the following
argument. At $z=0$, for $\alpha >0$ only the potential term contributes to the
energy density, whereas the gradient terms give no contribution. For $\alpha
=0$, instead, the potential and the quartic gradient term give exactly the same
contribution, as an immediate consequence of the Bogomolny nature of this 
solution. 
In the limit $\alpha \to 0$, this difference,
however, is of measure zero and does not influence the value of the energy, as
follows already from Figure (\ref{fig1}). 
We remark that these findings are in complete agreement with the general
discussion at the end of Section 3.

\begin{figure}[h!]
\begin{center}
\includegraphics[angle=-90.0,width=1.0 \textwidth]{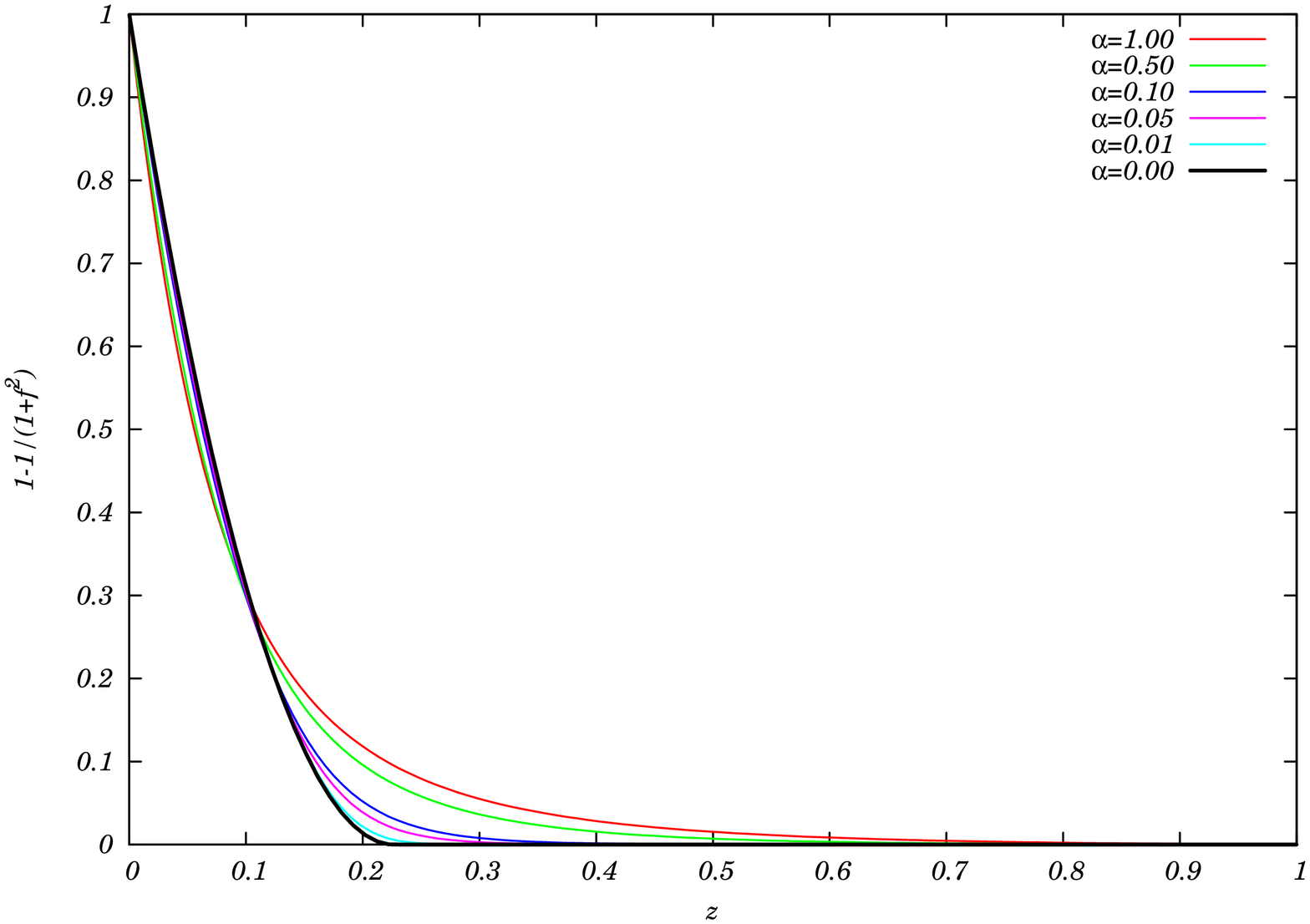}
\caption{Profile function $g$ in the full model, for different values of 
$\alpha$, and in the strongly coupled 
(BPS) model ($\alpha =0$) for Hopf charge $Q=m^2 = 1$, as a
function of $z$. 
The fixed parameter values are $R_0=5$, $\beta = 0.25$ and $\lambda =1$.
At $z=0$, the behaviour is linear for all values of
$\alpha$. }
\label{fig6}
\end{center}
\end{figure}

\begin{figure}[h!]
\begin{center}
\includegraphics[angle=-90.0,width=1.0 \textwidth]{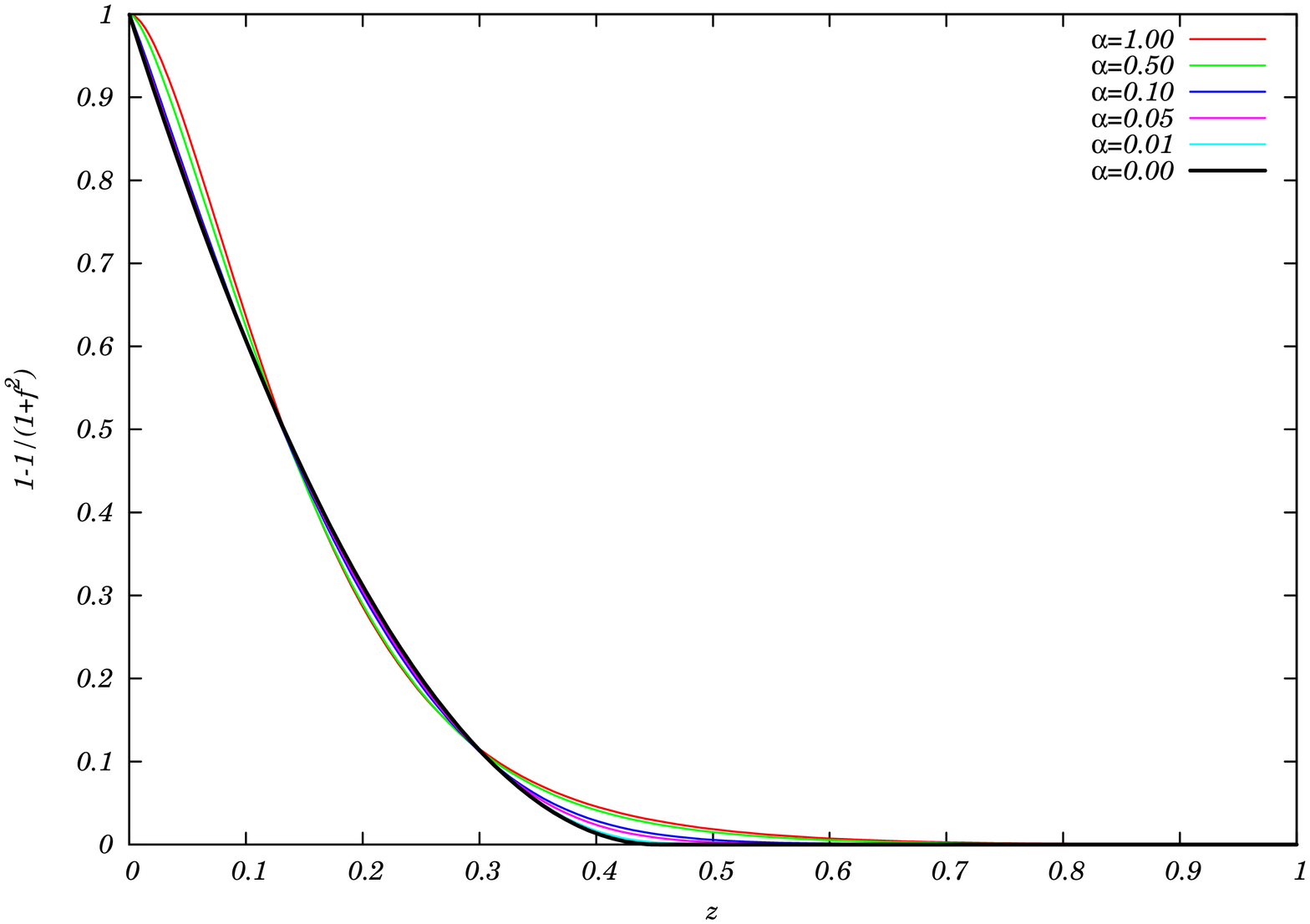}
\caption{Profile function $g$ in the full model, for different values of 
$\alpha$, and in the strongly coupled 
(BPS) model ($\alpha =0$) for Hopf charge $Q=m^2 = 4$, as a
function of $z$. 
The fixed parameter values are $R_0=5$, $\beta = 0.25$ and $\lambda =1$.
At $z=0$, the behaviour is linear for $\alpha =0$
and quadratic for
$\alpha >0$. }
\label{fig7}
\end{center}
\end{figure}

\begin{figure}[h!]
\begin{center}
\includegraphics[angle=-90.0,width=1.0 \textwidth]{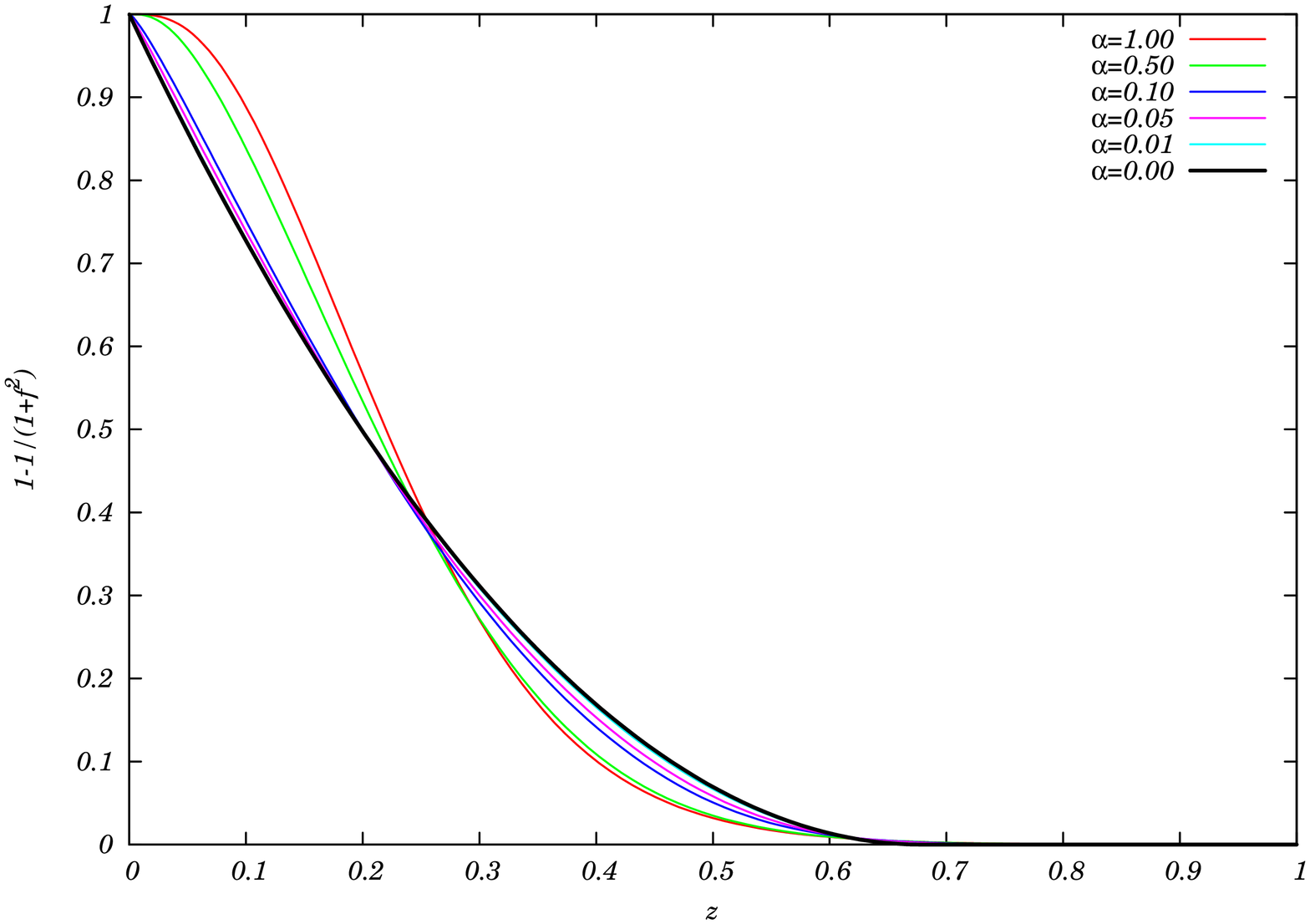}
\caption{Profile function $g$ in the full model, for different values of 
$\alpha$, and in the strongly coupled 
(BPS) model ($\alpha =0$) for Hopf charge $Q=m^2 = 9$ , as a
function of $z$. 
The fixed parameter values are $R_0=5$, $\beta = 0.25$ and $\lambda =1$.
At $z=0$, the behaviour is linear for $\alpha =0$
and cubic for
$\alpha >0$. }
\label{fig8}
\end{center}
\end{figure}

\begin{figure}[h!]
\begin{center}
\includegraphics[angle=-90.0,width=1.0 \textwidth]{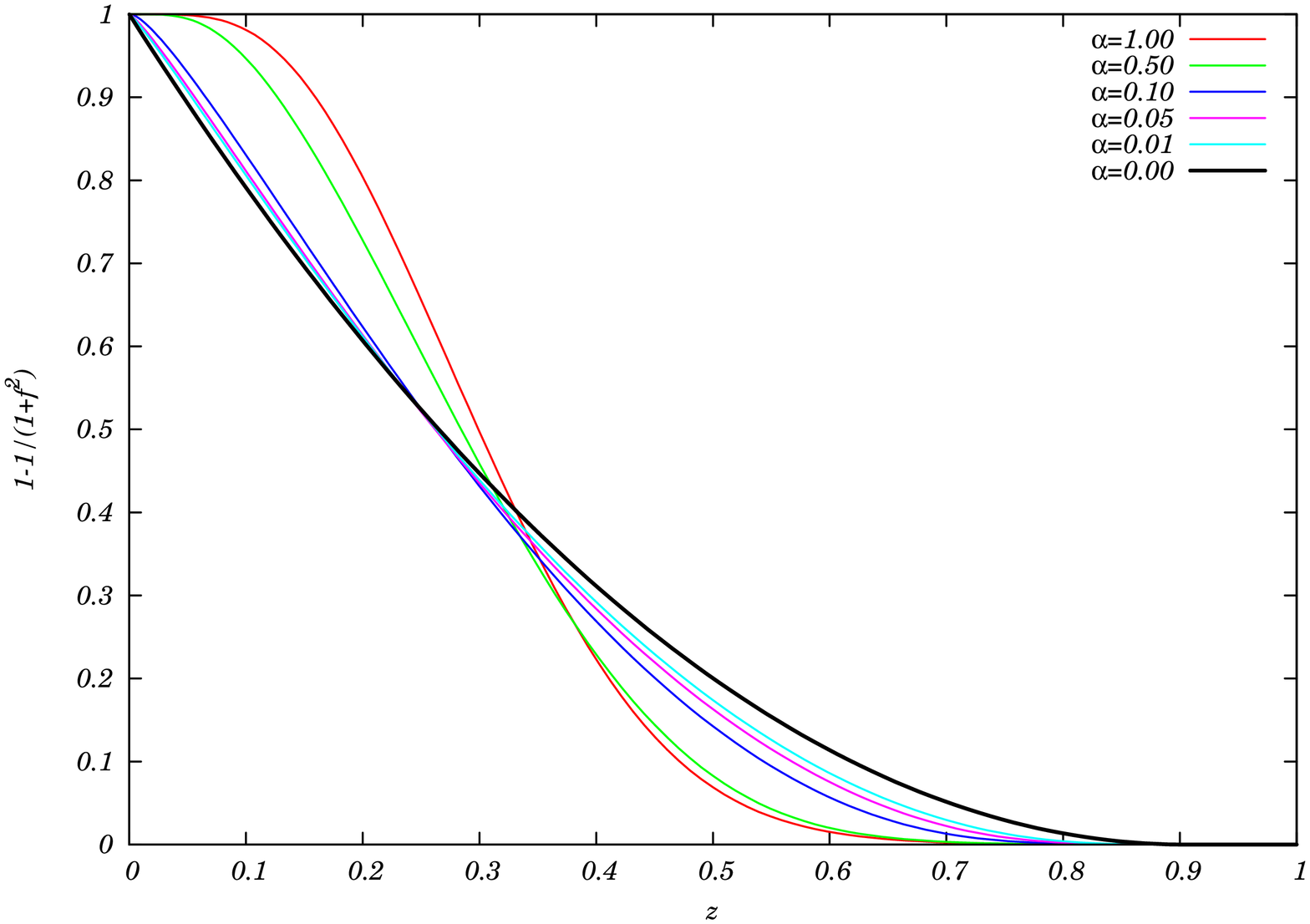}
\caption{Profile function $g$ in the full model, for different values of 
$\alpha$, and in the strongly coupled 
(BPS) model ($\alpha =0$) for Hopf charge $Q=m^2 = 16$ , as a
function of $z$. 
The fixed parameter values are $R_0=5$, $\beta = 0.25$ and $\lambda =1$.
At $z=0$, the behaviour is linear for $\alpha =0$
and quartic for
$\alpha >0$. }
\label{fig9}
\end{center}
\end{figure}

In Figures \ref{fig6} - \ref{fig9} 
we show the corresponding profile functions $g=1-(1/1+f^2)$,
for $m=1,2,3,4$. Again we find that the curves for small $\alpha$ approach the
curve for $\alpha =0$ uniformly in the case of $m=1$, whereas there remains a
small difference near $z=0$ for $|m|>1$. Indeed, for $\alpha =0$, $g$ behaves
linear, i.e., like $g \sim 1-c_1 z$ near $z=0$ for all $m$, whereas for
$\alpha >0$ $g$ behaves like $g\sim 1- c_m \, z^m$. 
\section{Conclusions}
It has been the main purpose of the present paper to investigate by means 
of analytical methods soliton solutions of the strongly coupled
Skyrme--Faddeev--Niemi model (with only a quartic kinetic term) with a
potential. We
explicitly constructed compact solutions, which are natural 
generalizations of the compact
solutions of the purely quartic baby Skyrme model which have first been
reported by Gisiger and Paranjape \cite{GP}, and further investigated 
recently \cite{comp baby}. As we wanted to present exact analytical 
solutions, we chose the base space (spacetime)    
$\mathbb{S}^3\times \mathbb{R}$ for finite energy solutions, because 
Minkowski spacetime does not offer sufficient symmetries to reduce the field
equations to ordinary differential equations. Only in the case of spinning
string-like solutions with a finite energy per length unit along the string
the symmetry reduction in Minkowski space is possible (Section 4).
For the case of $\mathbb{S}^3\times \mathbb{R}$ spacetime, we found two rather
different classes of finite energy soliton solutions, namely compactons (which
cover only a finite fraction of the three-sphere) on the one hand, and
non-compact solitons (which cover the full three-sphere) on the other
hand. Both classes of solutions are topological, but their energies are quite
different. The compacton energies behave like $E_c \sim R_0 |Q|^{1/2}$ (where
$R_0$ is the radius of the three-sphere, and $Q$ is the topological charge),
whereas the energies of the non-compact solitons behave like $E_s \sim R_0^3
|Q|$. Further, the compactons only exist up to a certain maximum value of the
topological charge, whereas the non-compact solitons start to exist from this
value onwards. The different behaviors of the energies in the compact and
non-compact case may be easily understood from the observation that the
compactons obey a Bogomolny equation, whereas the non-compact solitons obey a
``Bogomolny equation up to a constant''. Indeed, if for an energy density of
the type ${\cal E} = {\cal E}_4 + {\cal E}_0$ (here the subindices refer to
the power of first derivatives in each term) a Bogomolny equation holds, then
the energy density for solutions may be expressed like ${\cal E} \sim ({\cal
E}_4 {\cal E}_0)^{1/2}$. If we now take into account the scaling
dimensions ${\cal E}_4 \sim R_0^{-4}$, ${\cal E}_0 \sim R_0^0$ and $\int d^3 x
\sim R_0^3$, then the behaviour $E_c \sim R_0$ easily follows. 
Physically this means that the compacton solutions are localised near the
north pole of the three-sphere, and the localisation becomes more pronounced
for larger radii $R_0$. On the other hand, the energy density of the
non-compact solitons remains essentially delocalised and evenly distributed
over the whole three-sphere.
We remark that the behavior of the compacton energies $E_c \sim R_0
|Q|^{1/2}$ poses an apparent paradox, because it can be proven that already
the quartic part of the energy alone can be bound from below by $|Q|$, that
is, $E_4 \equiv \int d^3 x {\cal E}_4 \ge \alpha R_0^{-1} |Q|$, where $\alpha$
is an unspecified constant. The proof was given in \cite{SpSv1} for $R_0=1$,
but the generalization for arbitrary radius is trivial using the scaling
behavior of the corresponding terms. The apparent paradox is of course
resolved by the observation that compactons exist only for not too large
values of $|Q|$, such that the lower bound is compatible with the energies of
the explicit solutions. Finally, if the potential has more than one vacuum,
then compactons of the shell type exist, such that the field takes two
different vacuum values inside the inner and outside the outer compact shell
boundary. Except for their different shape, these compact shells behave quite
similarly to the compact balls in the one-vacuum case (e.g. the relation
between energy and topological charge or the linear growth of the energy with
the three-sphere radius is the same). 
\\
Further, we found that the strongly coupled model reproduces the properties of
the full model rather faithfully, at least on $\mathbb{S}^3$. Not only global
properties like the topological charge and the energy, but also issues like
the localized character (e.g., near the north pole) of a soliton are
common properties of solutions of the strongly coupled and the full model,
as we demonstrated in the numerical investigation in Section 5. 
We also found, however, that there exist some subtle, local differences 
for solutions
with a topological charge $|Q|>1$.
\\
To summarize, the inclusion of the potential term influences rather 
significantly qualitative as well 
as quantitative features of solitonic solutions: it modifies the 
energy-charge relation (especially for
small values of the topological charge) and it leads to ball or 
shell-type solitons for one or two
vacua potentials respectively.  
\\
One interesting question clearly is whether analogous properties 
(e.g.t the existence of compacton solutions with finite energy) can be 
observed in Minskowski space.  An exact calculation is probably
not possible in this case, but we think that we have found already some
indirect evidence for the existence of such solutions. The first argument is,
of course, the fact that they exist in one dimension lower (in the baby Skyrme
model). The second argument is related to the behaviour of our solutions for
large radius $R_0$. The compacton solutions are localized and, therefore,
their energies grow only moderately with $R_0$ (linearly in $R_0$). Further,
the allowed range of topological charges for compactons grows like the fourth
power of $R_0$. These are clear indications that compacton solutions might
also exist in Minkowski space. Certainly this question requires some further
investigation. If these compactons in Minkowski space exist, then an
interesting question is which energy-charge relation will result. Will the
energies grow like $E_c \sim |Q|^{1/2}$, like on the three-sphere, or will
they obey the three-quarter law $E_c \sim |Q|^{3/4}$, like for the full SFN
model without potential in Minkowski space? All we can say at the moment
is that an upper bound for the energy in flat space can be derived. The
derivation is completely analogous to the cases of the full SFN, Nicole or AFZ
models (the choice of trial functions which explicitly saturate the bound),
and also the result is the same, $E_c \le \alpha |Q|^{3/4}$, see
\cite{inequal}. The attempt to derive a lower bound, analogous to the
Vakulenku-Kapitanski bound for the SFN model, meets the same obstacles as for
the Nicole or AFZ models, see Appendix C of the second reference in
\cite{inequal}. 
\\    
Assuming for the moment the existence of compactons in Minkowski space, 
another interesting
proposal is to use the compacton solutions of the pure quartic model (with
potential) as a lowest order approximation to soliton solutions of the full
SFN model and try to approximate the full solitons by a kind of generalized
expansion. If such an approximate solution is possible, it would have several
advantages. 
\\
$\bullet$ The pure quartic model is much easier than the full
theory. In the case of the baby Skyrme model (both with old and new
potentials) one gets even solvable models (as long as the rotational
symmetry is assumed).
\\
$\bullet$ The lowest order solution is already a non-perturbative
configuration, i.e., a compacton, which captures the topological
properties of the full solution. Due to the compact nature of the
lowest order solution we have a kind of "localization" of the
topological properties in a finite volume.
\\
$\bullet$ One can easily construct multi-compacton solutions
which, if sufficiently separated, do not interact. They form something
which perhaps may be called {\it a fake Bogomolny sector} as they
are solutions of a first order equation (usually saturating a
corresponding energy-charge inequality) and may form multi-soliton
noninteracting complexes.
\\
Of course, it remains to be seen whether such an approximate solution is
possible at all. What can be said so far is that in the simpler case of a
scalar field theory with a potential which is smooth if a certain parameter 
$\mu$ is
non-zero and approaches a V-shaped potential in the $\mu \to 0$ limit, 
then the compacton is the $\mu \to 0$ limit of the non-compact soliton,
see \cite{lis}. Similarly, as was shown in the last section, the Hopf 
compactons of the strongly 
coupled model approximate the solitons of the full SFN theory, at least 
on $\mathbb{S}^3$ space. 

\section*{Acknowledgements}

C.A. and J.S.-G. thank MCyT (Spain) and FEDER (FPA2005-01963), and
support from Xunta de Galicia (grant PGIDIT06PXIB296182PR and
Conselleria de Educacion). A.W. acknowledges support from the
Ministry of Science and Higher Education of Poland grant N N202
126735 (2008-2010). 
J.S.-G. visited the Institute of Physics, Jagiellonian University,
Krak\'ow, during the early stages of this work. He wants to thank the
Institute for its hospitality and a very stimulating work environment. 
Further, A.W. thanks M. Speight, J. J\"{a}ykk\"{a} and P. Bizo\'{n} for 
helpful discussions.

\end{document}